\documentclass[prd,preprint,superscriptaddress,amsmath,amssymb]{revtex4}
\usepackage{graphicx,color}% Include figure files
\usepackage{slashed}

\begin{document}

%%%%%%%%%
\title{Light gauge boson in rare $K$ decay}
\preprint{KIAS-P16055}
\author{Chuan-Hung Chen}
\email{physchen@mail.ncku.edu.tw}
\affiliation{Department of Physics, National Cheng-Kung University, Tainan 70101, Taiwan}

\author{Takaaki Nomura}
\email{nomura@kias.re.kr}
\affiliation{School of Physics, KIAS, Seoul 130-722, Korea}

\date{\today}

\begin{abstract}

We study the production of a light gauge boson in $K^- \to \pi^- X$ decay, where the associated new charge current is not conserved. 
 It is found that the process can be generated by the tree-level $W$-boson annihilation and loop-induced $s\to dX$.  We find  that it strongly depends on the $SU(3)$ limit or the unique gauge coupling to the quarks, whether the decay amplitude of  $K^-\to \pi^- X$  in the $W$-boson annihilation is suppressed by $m^2_X \epsilon_X \cdot p_K$; however, no such suppression is found via the loop-induced $s\to d X$.  The constraints on the relevant couplings are studied. 
 
\end{abstract}
\maketitle

A light gauge boson $X$ has been studied widely for various reasons~\cite{Gninenko:2001hx,Fayet:2007ua,ArkaniHamed:2008qn,Pospelov:2008zw,Davoudiasl:2014kua,Petraki:2014uza,Lee:2016ief,Araki:2015mya,Baek:2015fea,Ko:2016uft}. In particular, a mass of around 17 MeV boson with more than $5\sigma$ significance is indicated by the  measurement of $e^+ e^-$ angular correlations  in the $^8Be$ transition~\cite{Krasznahorkay:2015iga}, where the implications of this light gauge boson are  investigated~\cite{Feng:2016jff,Gu:2016ege,Chen:2016dhm,Liang:2016ffe}. 

However, two conclusions associated with the $X$ emission in  rare $K$ decay appear in the literature, where some authors~\cite{Fayet:1980rr,Suzuki:1986kt}  concluded that the longitudinal component in the $K^-\to \pi^- X$ decay is enhanced by $\epsilon_X \cdot p_K \sim m^2_K/m_X$, but others \cite{Pospelov:2008zw,Davoudiasl:2014kua,Aliev:1988ks} showed that this decay amplitude should be suppressed by $ \epsilon_X\cdot p_K m^2_X/m^2_K \sim m_X$. 

In general,  the decay amplitude for the  $K^-\to \pi^- X$ process can be written as $A_\lambda=\langle X \pi^- | H_{I} | K^- \rangle = M_\mu \epsilon^\mu_X(k,\lambda)$, where $H_I$ is the involved interaction; $M_\mu$ is the transition matrix element for $K^+\to \pi^+$,  and $\epsilon^\mu_X(\lambda)$ denotes the $X$ polarization vector  with the momentum $k$ and helicity $\lambda$. Thus, the spin-average amplitude square can be expressed as:
  \begin{equation}
  \sum_\lambda |A_\lambda|^2 = M_\mu M_\nu \left(- g^{\mu \nu} + \frac{k^\mu k^\nu}{m^2_X} \right)\,.
  \end{equation}
 If the charge, which is associated with gauge symmetry for the $X$-gauge boson, is conserved, following the current conservation $k^\mu M_\mu =0$, it can be  seen that the term $k^\mu k^\nu/m^2_X$ vanishes. Clearly, the $1/m_X$ enhancement for a light gauge boson is associated with the charge nonconservation, i.e.,  $k^\mu M_\mu \neq 0$. Accordingly,  the $1/m_X$ factor indeed is suppressed when the associated current is conserved, such as the case of a dark photon that mixes with the photon through the kinetic term~\cite{Holdom:1985ag, Jaeckel:2012yz}. Furthermore,  due to gauge invariance, the decay amplitude for $K^+\to \pi^+ X$  in such cases should vanish at the tree level according to  chiral perturbation theory~\cite{Ecker:1987qi};  the main contributions then are from the loop effects.  A detailed analysis about the dark-photon case can be found in Refs.~\cite{Pospelov:2008zw,Davoudiasl:2014kua}.
 
 To obtain a  further understanding the properties of $K^-\to \pi^- X$ in the model with charge nonconservation,  in this study,  we analyze  this issue by exploring the situations with and without the $SU(3)$ limit and unique gauge coupling when the $K^-\to \pi^-$ transition arises from the $W$-boson annihilation. In addition, we also study the contributions from  the loop-induced flavor-changing neutral current (FCNC) process $s\to dX$.

Since our purpose is to investigate the properties of a  light gauge boson emission from the $K^-\to \pi^-$ transition,  we do not focus on a specific gauge model. Instead, we study  a case in which  the  $X$-boson vectorially couples to the standard model (SM) quarks and the interactions are dictated by:
\begin{equation}
{\cal L}_{qq'X} = g_{qq'} \bar q \gamma_\mu q' X^\mu\,. \label{eq:LX}
\end{equation}
In general, the  couplings $g_{qq'}$ are flavor-dependent, and  the FCNCs at the tree level are then  induced. Since we have little knowledge on the flavor mixings, the associated FCNC parameters are completely free. We thus skip discussions on the tree-level FCNC effects in this work by assuming that they are small, or that they can always be  constrained by low energy physics. In the following analysis, we focus on couplings with $q=q'$ by writing $g_q \equiv g_{qq}$ for simplicity.

%Equation~(\ref{eq:LX}) can be applied to dark photon models if  $g_q$ for the same type of quarks are the same~\cite{Holdom:%1985ag, Jaeckel:2012yz}. 
% Hence, we just concentrate on the contributions from Eq.~(\ref{eq:LX}).  

In order to demonstrate the characteristics of  the $K^-\to \pi^- X$ decay,  we analyze the hadronic effects with leading-twist parton  distribution amplitudes (DAs) for the $\pi$ and $K$ mesons. As usual, 
the twist-2 DA of a pseudoscalar meson is defined by \cite{Braun:1989iv,Ball:1998je}:
  \begin{align}
 \langle 0 | \bar q'(x) \gamma_5 \gamma_\mu q(-x) | P(p) \rangle & = -i f_P p_\mu \int^1_0 du e^{i\xi  p\cdot x} \phi_P (u)\,,
   \end{align}
   where  $\xi=2 u-1$, $\int^1_0 du\, \phi_P(u) =1$, and  $f_P$ is the decay constant of a meson $P$.  The  DA can be expanded by  Gegenbauer polynomials as:
   \begin{equation}
   \phi_P (u) = 6 u (1-u)  \left(1 + \sum_{i=1} a^P_i(\mu) C^{3/2}_i (2u-1) \right)\,, \label{eq:wf}
   \end{equation}
   where the Gegenbauer moments $a^P_i$ for  the $\pi$ and $K$ mesons at $\mu=1$ GeV are $a^\pi_{2i+1}=0$, $a^\pi_2 = 0.44$, $a^\pi_4=0.25$, $a^K_1=0.17$, and $a^K_2 =0.2$~\cite{Braun:1989iv,Ball:1998je,Ball:1998tj}. It can be seen that due to breaking of the $SU(3)$, the odd moments in the $K$ meson do not vanish. 
   To calculate the $X$ emission from the $K$ and $\pi$ mesons, we adopt the spin structure for incoming meson  as~\cite{Chen:2001pr,Chen:2006vs}:
  \begin{align}
   \langle 0 | \bar q_{1\beta}(0) q_{2\alpha}(z) | P(p) \rangle =  \frac{-i f_P}{4N_c} \int^1_0 dx e^{-i x p\cdot z} [\slashed{p} \gamma_5]_{\alpha\beta} \phi_P(x) \,, \label{eq:DA}
   %
%    \langle P(p) | \bar q_{2\beta}(z) q_{1\alpha}(0) |0 \rangle =  \frac{-i}{4N_c} \int^1_0 dx e^{i x p\cdot z} \gamma_5 \not{p} \phi_P(x)
  \end{align}
where  $N_c=3$ is the number of colors, and the spin structure  for outgoing meson can be obtained by  using $\gamma_5\slashed{p}$ instead of $\slashed{p} \gamma_5$. According to Eq.~(\ref{eq:DA}), the $f_P$ can be obtained  by taking the trace in  spinor space as:
 \begin{eqnarray}
 \langle 0 | \bar q_2 \gamma_\mu \gamma_5 q_1 | P(p) \rangle = -i \frac{N_c f_P}{4N_c}\int^1_0 Tr(\gamma_\mu \gamma_5 \slashed{p}\gamma_5) \phi_P = i f_P p_\mu \,,
 \end{eqnarray}
where the $N_c$   in the numerator is from the sum of color charges of  quark line $\bar q_{2\alpha}[...] q_{1\alpha}$.

%%%%%%%%%%%%%%%%%%%%%%%%%%%%%%%%%%%%%%%%%%%%%%%%%%%%%%%%%%%%%%%%%%
\begin{figure}[hpbt] 
\includegraphics*[width=75mm]{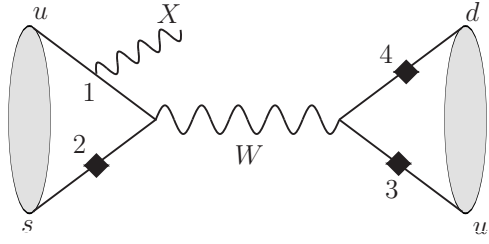} 
\caption{ Flavor diagrams for the tree-level $K^-\to \pi^- X$ decay, where the square boxes and numbers denote the possible places to emit the  $X$-boson.}
\label{fig:KpiX}
\end{figure}
%%%%%%%%%%%%%%%%%%%%%%%%%%%%%%%%%%%%%%%%%%%%%%%%%%%%%%%%%%

With the couplings in Eq.~(\ref{eq:LX}), the $K^-\to \pi^- X$ decay can arise from the tree and one-loop diagrams. Since the hadronic effects  from the tree level  are more copious than those from one-loop, we first discuss the tree contributions in detail. The flavor diagrams for the $K^- \to \pi^- X$ decay are shown in Fig.~\ref{fig:KpiX}, where the square boxes and numbers denote the possible places to emit the  $X$-boson. According to  Eq.~(\ref{eq:DA}), the decay amplitude for the $X$-boson emitting from place-1 can be derived as:
%%%%
 \begin{align}
  i A_1 
%  &  \approx  \langle \pi^-| \frac{-i g}{\sqrt{2}}  V^*_{ud} \bar d \gamma_\nu P_L u |0\rangle  \frac{ig^{\nu \mu}}{m^2_W} 
%  \langle X | \bar u \frac{-i g}{\sqrt{2}} V_{us} \gamma_\mu P_L s  | K^- \rangle \,, \nonumber \\
  & \approx \frac{-i g^2}{2 m^2_W } V^*_{ud}V_{us}  \left( \frac{i f_\pi p^\mu_{\pi}}{2} \right)  \nonumber \\
 &  \times \int^1_0 dx \,Tr\left( i g_u \slashed{\epsilon}_X \frac{i}{-\slashed{p}_u + \slashed{p}_X - m_u} \gamma_\mu P_L \slashed {p}_K \gamma_5 \right)\frac{-i N_c f_K \phi_K(x) }{4 N_c}\,.
 \end{align}
%%%%%
It is found that the decay amplitude for the $X$-boson emitting  from place-$j$ can be formulated as:
 \begin{equation}
 A_j \approx  g_j C_{K\pi} \epsilon_X \cdot p_K \int^1_0 dx R_j( p_K, p_\pi, x) \phi_j(x) \,, \label{eq:A_j}
 \end{equation}
where  $C_{K\pi} = G_F/(2\sqrt{2}) V^*_{ud} V_{us} f_K f_\pi $, $g_{1,3} =g_u$, $g_2=g_s$, $g_4=g_d$, $\phi_{1,2}=\phi_K$, $\phi_{3,4} =\phi_\pi $, and $R_j$ are defined as:
  \begin{align}
  & R_1 (p_K, p_\pi, x)= \frac{-2x p^2_\pi -2(1-x)m^2_X }{m^2_X -2 x p_X \cdot p_K} \,, ~
   R_2(p_K,p_\pi,x)=-R_1(p_K,p_\pi,1-x)\,,  \nonumber \\
   & R_3(p_K, p_\pi, x)=R_1(-p_\pi, p_K, x)\,, ~ R_4(p_K,p_\pi, x)=-R_1(-p_\pi, p_K, 1-x). \label{eq:R_j}
  \end{align}
 In order to understand  the $SU(3)$ limit and  $g_q$ dependence of $A_j$, we study the cases by requiring an exact/partial $SU(3)$ limit and different/same $g_q$. 

  \underline{(I) SU(3) limit: $m_K = m_\pi \equiv m_P$, $\phi_K = \phi_p \equiv \phi_P$}: \\
% %%%
  \noindent Since $f_K$ and $f_\pi$ are  the multiplier factors,  they are irrelevant to the discussions of the $SU(3)$ limit; therefore, we do not need to set $f_K=f_\pi$.  Due to the kinematics, $p_K = p_\pi + p_X$ has to be satisfied; thus, we need to leave the factors $p_X\cdot p_K$ and $p_X\cdot p_\pi$, which appear in the denominators of Eq.~(\ref{eq:R_j}), alone. From Eqs.~(\ref{eq:A_j}) and (\ref{eq:R_j}), we then get:
   \begin{align}
   A_1 + A_3 = -   g_u C_{k\pi}  \epsilon_X\cdot p_K  m^2_X \int^1_0 dx \phi_P(x) \frac{ 4 (1-x)\left( x m^2_P + (1-x) m^2_X \right)}{(m^2_X + 2x p_X\cdot p_\pi) (m^2_X -2 x p_X\cdot p_K)}\,, \label{eq:A1pA3}
   \end{align}
and  $A_2+A_4 = -(g_d+g_s)/(2 g_u )(A_1 + A_3)$. We consider $A_1+A_3$ because $A_1$ and $A_3$ involve the same gauge coupling $g_u$. One can  alternatively use  $A_1+A_2 (A_3+A_4)$ according to  the convenience. It  is clearly seen that  the decay amplitude for  the $K^-\to \pi^- X$  is proportional to $m^2_X$. This result matches the conclusions given in two earlier works~\cite{Pospelov:2008zw,Aliev:1988ks}. We note that when calculating $A_2+A_4$, we have used the property $\phi_P(x)=\phi_P(1-x)$, where this condition is suitable for the $\pi$ meson, and it is violated in the $K$ meson due to the breaking of $SU(3)$. As a result,    the leading-twist contributions  can not lead to an interesting result in the $SU(3)$ limit. Furthermore,  if we further set $g_s+g_d=2g_u$,  it can be found that  $\sum A_i =0$.   

\underline{(II) Partial SU(3) limit: $\phi_K = \phi_\pi \equiv\phi_P$}\\
% %%%
\noindent  When we release the condition $m_K= m_\pi$, the first term in the numerator of Eq.~(\ref{eq:A1pA3}) inside the integral becomes $m^2_X (m^2_K + m^2_\pi)(1-x) -(m^2_K -m^2_\pi)^2 x$, and the denominator is $m^4_X (1-x)^2 -(m^2_K -m^2_\pi)^2 x^2$. With $m_X = 0$, we find:
 \begin{align}
 A_1 + A_3 & = -2 g_u C_{k\pi}  \epsilon_X\cdot p_K\,, \nonumber \\
 A_2+A_4 & = 2 \frac{ g_d m^2_K - g_s m^2_\pi}{m^2_K -m^2_\pi}  C_{K\pi} \epsilon_K \cdot p_K \,,
 \end{align} 
where the $x$ dependence of  the numerator and denominator in the integral is cancelled,  and $\int^1_0 \phi_P(x) dx =1$ is applied. By this analysis, it is clear that when we put back the $SU(3)$ breaking effect with $m_K \neq m_\pi$, the decay amplitude is not proportional to $m^2_X$ anymore. This result confirms the conclusions given in two earlier studies~\cite{Fayet:1980rr,Suzuki:1986kt}. Furthermore, if we take all gauge couplings to be the same and $m^2_X \neq 0$, we find  that $\sum_i A_i = 0$ is still satisfied.  We can understand the cancellations from another viewpoint: by using  $\phi_P (x)=\phi_P(1-x)$, from Eqs.~(\ref{eq:A_j}) and (\ref{eq:R_j}) it can be easily found that  $A_1+A_2 =0$ for $g_u=g_s$ and  $A_3 + A_4 =0$ for $g_u=g_d$. 
%That is, when  $\phi_K = \phi_\pi$ is adopted and only one gauge coupling to the quarks is involved, the decay amplitude for the 
%$K\to \pi X$ vanishes.
%, although it is not proportional to $m^2_X$. 

 \underline{(III) $SU(3)$ breaking }\\
 \noindent With a partial $SU(3)$ limit, which is conditioned by $\phi_K = \phi_\pi$,  it can be seen that the decay amplitude for the $K^-\to \pi^- X$ is not suppressed by $m^2_X$; however, it diminishes when $g_s\sim g_d \sim g_u$. It is intriguing to see whether  the cancellations 
work or not when the $SU(3)$ breaking effects are taken into account in the DA of the $K$ meson. From Eqs.~(\ref{eq:A_j}) and (\ref{eq:R_j}),  we can easily get the result $A_4 = - g_d/g_u A_3$ when the $g_u$ and $x$  in $A_3$ are replaced by the $g_d$ and $1-x$. The connection between $A_3$ and $A_4$ is based on the property $\phi_\pi(x) = \phi_\pi (1-x)$, where the odd Gegenbauer moments vanish. That is, if $g_d=g_u$, the cancellation between  $A_3$ and $A_4$ still works. Unlike the DA of pion, $\phi_K(x) \neq \phi_K(1-x)$ due to nonvanishing odd Gegenbauer moments, e.g., $a^K_1 =0.17$. Hence, we have: 
 \begin{equation}
 A_1+A_2 =  C_{K\pi} \epsilon_X\cdot p_K\int^1_0 R_1 (p_K , p_\pi,x) \left[ g_u \phi_K (x)- g_s \phi_K(1-x) \right] \label{eq:A1pA2}
 \end{equation}
  In order to examine Eq.~(\ref{eq:A1pA2}),  we simplify the analysis by taking the limit $m_X\to 0$. Thus, with $g_s=g_u$,  we find:
   \begin{equation}
   A_1 + A_2 \propto a^K_1 \int^1_0 x^2 (1-x) (1-2x) =0\,.
   \end{equation}
  According to the result, it can be seen that the decay amplitude for the $K^-\to \pi^- X$ with $g_u=g_s=g_d$ is $\sum_i A_i \propto a^K_1 f(m^2_X)$, where the function  $f(m^2_X)$ is from the integration in $x$ and only depends on the $m^2_X$. We then conclude that if the $X$ gauge couplings to the quarks are the same, the decay amplitude of the $K^-\to \pi^- X$ process from the leading-twist DA   is suppressed by $m_X$.   To illustrate  the relative magnitude under the $SU(3)$ assumptions,  we show the numerical values with some chosen values of couplings and $m_X$ in Table~\ref{tab:SU3}, where the function $M_{m_X}$ is defined as:
 \begin{align}
%  M(g_u,g_s,g_d)_{m_X} &= \int ^1_0  dx \left[ g_u \left(R_1(p_K,p_\pi,x) \phi_{K}(x) + R_4(p_K,p_\pi,x)\phi_\pi (x) \right) \right. 
%\nonumber \\
 % &  \left. + g_s R_2(p_K,p_\pi,x)\phi_K(x) + g_d R_3(p_K,p_\pi,x) \phi_\pi(x)\right]
 M(g_u,g_s,g_d)_{m_X}= \frac{1}{C_{K\pi} \epsilon_X\cdot p_K} \sum^4_{j=1} A_j\,. \label{eq:MX}
  \end{align} 
  $SU(3)_{I,II,III}$ denote the cases for the $SU(3)$ limit, the partial $SU(3)$ limit, and the breaking of $SU(3)$. 
    
   \begin{table}[hpbt]
 %\begin{ruledtabular}
\begin{tabular}{c|ccc} 
\hline \hline
$M(g_u,g_s,g_d)_{m_X}$   & $~~SU(3)_I~~$ &  $~~SU(3)_{II}~~$ &  $~~SU(3)_{III}~~$  \\ \hline
$M(g,g,g)_{10 MeV}$  & 0 & 0 & $-9\cdot 10^{-4} g$ \\ \hline
$M(2g,g,g)_{ 10 MeV}$ & 0.35$g$ & 2.35$g$ & 2.35$g$  \\ \hline
$M(g,g,g)_{100 MeV}$ & 0 & 0 & $-0.1 g$ \\ \hline
$M(2g,g,g)_{ 100 MeV}$ &  0.79$g$ & 2.58$g$ & 2.43$g$ \\ \hline
\end{tabular}
\caption{ Magnitude with some chosen values of couplings and  $m_X$ under the $SU(3)$ assumptions, where $M(g_u,g_s,g_d)_{m_X}$ is defined in Eq.~(\ref{eq:MX}). }
\label{tab:SU3}
%\end{ruledtabular}
\end{table}

  It is of interest to examine the $X$ emission from the $W$-boson propagator shown in Fig.~\ref{fig:KpiX}. To estimate the contribution,  we parametrize the Lorentz covariant gauge coupling $WWX$ to be:
  \begin{equation}
  {\cal L} \supset g_{WWX} \left[g_{\alpha\beta}(p_- - p_+)_\mu + g_{\beta \mu}(p_+ - p_X)_\alpha 
  + g_{\mu \alpha}(p_X - p_-)_\beta \right] W^{-\alpha} W^{+\beta}X^\mu\,,
  \end{equation}
 where $g_{WWX}$ is the trilinear gauge coupling; $p_{-}$, $p_+$, and $p_X$ are the momenta of the $W^-$, $W^+$, and $X$ gauge bosons, respectively, and the momenta are chosen to flow into the vertex.  Accordingly, the decay amplitude for the $K^-\to \pi^- X$ via the trilinear coupling can be obtained as:
 \begin{equation}
 A_{WWX} =  V^*_{ud} V_{us} \frac{G_F f_K f_\pi}{\sqrt{2}}  \frac{m^2_X}{m^2_W} g_{WWX} \epsilon_X\cdot p_K\,.
 \end{equation}
 It is clear that the contribution from the coupling $WWX$ is suppressed by $m_X$. This result is nothing to do with the $SU(3)$ limit.

\begin{figure}[hpbt] 
\includegraphics*[width=80mm]{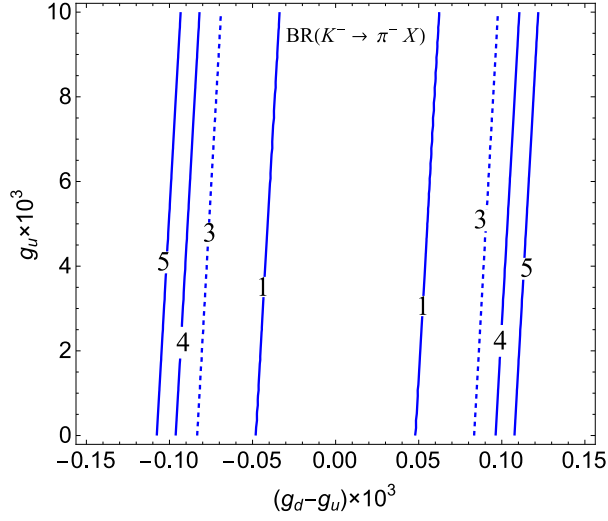} 
\caption{Contours for $BR(K^-\to \pi^- X)$ in units of $10^{-7}$ as a function of $g_u$ and $g_d-g_u$ in units of $10^{-3}$, where the numbers on the lines denote the values of $BR(K^-\to \pi^- X)$, and dashed lines are the central value of $BR^{\rm exp}(K^-\to \pi^- e^+ e^-)$.  }
\label{fig:WA}
\end{figure}

 Without the $SU(3)$ limit, it can be seen from Eq.~(\ref{eq:R_j}) that the numerators of $R_{1,2}$ are related to $m^2_\pi$ and that those of $R_{3,4}$ are associated with $m^2_K$. Since $m_K$ is around 3.5 times larger than $m_\pi$, numerically, the values of $A_{1,2}$ are one order of magnitude smaller than those of $A_{3,4}$. As a result, the $BR(K^-\to \pi^- X)$ is sensitive to the difference between $g_d$ and $g_u$.  To present the numerical analysis, we adopt $g_u$ and $g_d - g_u$ as the free parameters and set $g_s = g_d$. We show the contours for $BR(K^-\to \pi^- X)$ (in units of $10^{-7}$) as a function of $g_u$ and $g_d -g_u$ (in units of $10^{-3}$) in Fig.~\ref{fig:WA}, where we have used $f_K=0.16$ GeV and $f_\pi=0.13$ GeV, and the numbers on the lines denote the values of $BR(K^-\to \pi^- X)$. With the assumption of  $BR(X\to e^+ e^-)\sim 1$, we obtain $BR(K^-\to \pi^- e^+ e^-)\approx BR(K^-\to \pi^- X)$, where the current measurement is $BR^{\rm exp}(K^- \to \pi^- e^+ e^-) = (3.00 \pm 0.09)\times 10^{-7}$~\cite{PDG}. Therefore, 
  the dashed lines in the plot can be regarded as the central value of  the experimental measurement. From the figure it can be seen that $|g_d-g_u|$ cannot be larger than $10^{-4}$.

 %%%%%%%%%%%%%%%%%%%%%%%%%%%%%%%%%%%%%%%%%%%%%%%%%%%%%%%%%%%%%%%%%%
%%%%%%%%%%%%%%%%%%%%%%%%%%%%%%%%%%%%%%%%%%%%%%%%%%%%%%%%%%

\begin{figure}[hpbt] 
\includegraphics*[width=75mm]{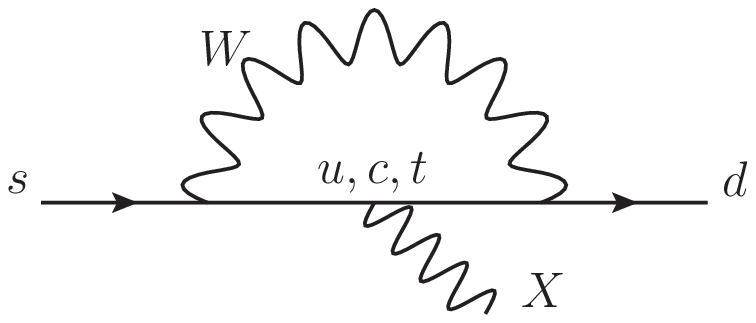} 
\caption{One-loop Feynman diagram for $s\to d X$.  }
\label{fig:loop}
\end{figure}

In addition to  the tree-level $W$-boson annihilation, the FCNC coupling $sd X$, which is induced from one-loop and is depicted in Fig.~\ref{fig:loop}, can also contribute to the $K^-\to \pi^- X$ process. According to the interactions in Eq.~(\ref{eq:LX}), the effective coupling of  $sd X$ from each up-type quark loop can be derived as:
\begin{align}
A_{q} & =C_q  \bar d \gamma_\mu (1-\gamma_5) s X^\mu\,,  \label{eq:sdX}\\
 C_q & = V_{qs} V^*_{qd}  \frac{4 G_F}{\sqrt{2}}\frac{g_q m^2_q}{(4\pi)^2} 
 I_q\left( \frac{m^2_q}{m^2_W}\right)\,, \nonumber \\
 I_q (r) & \approx \int^1_0 dy \frac{y}{1-(1-r) y} \,, \nonumber 
\end{align}
where we have dropped the small effects from $m_{s,d}$; the factor $m^2_q$ is from the  mass inserted twice in $q$-quark propagator,  and $I_q(r)$ is the loop integral. From Eq.~(\ref{eq:sdX}),   the associated Cabibbo-Kobayashi-Maskawa (CKM) matrix elements  for top-quark loop  are $V^*_{td} V_{ts}$, and  due to the enhancement of $m^2_t$, its contribution is comparable to that from the charm-quark, in which the essential factor is  $m^2_c V^*_{cd} V_{cs}$.  The contribution  from the $u$-quark loop can be ignored because of the $m^2_u$ suppression.  Additionally, it can be clearly seen that although the $X$ couplings to quarks are vector-like, the induced coupling $sdX$ indeed is chiral.

%%%%%%%%%%%%%%%%%%%%%%%%%%%%%%%%%%%%%%%%%%%%%%%%%%%%%%%%%%
%%%%%%%%%%%%%%%%%%%%%%%%%%%%%%%%%%%%%%%%%%%%%%%%%%%%%%%%%%

\begin{figure}[hpbt] 
\includegraphics*[width=80mm]{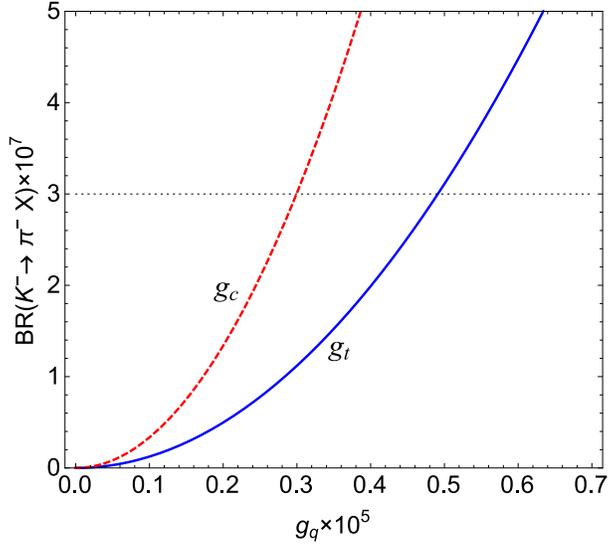} 
\caption{Loop-induced $BR(K^-\to \pi^- X)$ as a function of $g_q$  (in units of $10^{-5}$), where the solid and dashed lines denote the contributions from top- and charm-quark loops, respectively. The horizontal dotted line is the central value of $BR^{\rm exp}(K^-\to \pi^- e^+ e^-)$.   }
\label{fig:BR_loop}
\end{figure}

Unlike the $W$-boson annihilation shown in Fig.~\ref{fig:KpiX}, the dominant  hadronic effect for the $K^-\to \pi^- X$ decay from $sdX$ interaction is formulated by~\cite{Carrasco:2016kpy}:
 \begin{eqnarray}
 \langle \pi^- | \bar d \slashed{\epsilon}_X s | K^-\rangle = 2 \epsilon_X\cdot p_K f_+(q^2) \,,
 \end{eqnarray}  
where $f_+ \approx 0.971$ at $q^2=0$. As a result, the corresponding  transition amplitude  and BR are respectively given by:
 \begin{align}
 \langle \pi^- X | A_q | K^- \rangle & = 2 C_q f_+(m^2_X) \epsilon_X \cdot p_K \,, \nonumber \\
 BR(K^-\to \pi^- X)  & = \frac{|C_q f_+(m^2_X)|^2}{2\pi} \frac{\tau_{K^+} |\vec{p}_X|^3}{m^2_X}\,. \label{eq:KpiX_loop}
 \end{align}
It can be seen that  Eq.~(\ref{eq:KpiX_loop}) is not suppressed by $m_X$ but rather is enhanced by $1/m_X$.   This result is consistent with the dark $Z$ model in Ref.~\cite{Davoudiasl:2014kua} if we set $g_q \sim \delta\, m_{X}/m_Z$. To show the bounds on the gauge couplings $g_t$ and $g_c$ independently, we present the numerical values of Eq.~(\ref{eq:KpiX_loop}) as a function of $g_t$ and $g_c$ in Fig.~\ref{fig:BR_loop}, where $m_X=17$ MeV is used, the solid (dashed) line stands for the result of  $g_t (g_c)$ in units of $10^{-5}$, and the horizontal dotted line is the central value of $BR^{\rm exp}(K^-\to \pi^- e^+ e^-)$. With $BR(X\to e^+ e^-)\sim 1$, the bounds from the penguin diagrams are stronger than those from the $W$-boson annihilations. Therefore, we confirm the conclusion given in a previous work~\cite{Suzuki:1986kt}, where   the effective coupling arising from the penguin diagrams obtains a stricter bound. We note that  the $X$-boson emitting from the $W$ propagator shown in Fig.~\ref{fig:loop} can also contribute to $K^-\to \pi^- X$; however, due to the $m^2_K/m^2_W$ suppression, we neglect its contribution in the numerical analysis.

In summary, a light gauge boson predominantly decaying to $e^+ e^-$ in rare $K$ decay is studied. The process $K^-\to \pi^- X\to \pi^- e^+ e^-$  can be generated from both tree and penguin diagrams. It is found that the decay amplitude for $K^-\to \pi^- X$ from $W$-boson annihilation can be directly proportional to $m^2_X\epsilon_X \cdot p_K$ when  the $SU(3)$ limit is applied, or when the gauge couplings satisfy $g_u=g_d=g_s$. When these conditions are relaxed, it is found that  if $g_u$ is of the ${\cal O}(10^{-3})$, $|g_d-g_u|$ has to be less than $10^{-4}$. By contrast, the $m^2_X$ suppression is not found in the loop-induced $K^-\to \pi^- X$ process. We show that  the loop effects indeed  produce more severe bounds on the gauge couplings $g_t$ and $g_c$. Our results are consistent with the conclusions given in two previous studies~\cite{Fayet:1980rr,Suzuki:1986kt}.

 %%%%%%%%%%%%%%%%%%%%%%%%%%%%%%%%%%%%%%%%%%%%%%%%%%%%%%%%%%%%%%%%%%
%%%%%%%%%%%%%%%%%%%%%%%%%%%%%%%%%%%%%%%%%%%%%%%%%%%%%%%%%%

\section*{Acknowledgments}

This work was partially supported by the Ministry of Science and Technology of Taiwan
R.O.C.,  under grant MOST-103-2112-M-006-004-MY3 (CHC). 

%%%%%%%%%%%%%%%%%%%%%%%%%%%%%%%%%%%

\end{document}